\documentclass[aps,prb,twocolumn,showpacs,superscriptaddress]{revtex4}
\usepackage{graphicx}
\usepackage{dcolumn}
\usepackage{bm}
\usepackage{amssymb}
\usepackage{amsmath}
\usepackage{epsfig}

\begin{document}

\title{Universal Quantum Degeneracy Point for Superconducting Qubits}

\author{X.-H. Deng}
\author{Y. Hu}
\author{L. Tian}
\email{ltian@ucmerced.edu}
\affiliation{5200 North Lake Road, University of California, Merced, CA 95343, USA}

\begin{abstract}
The quantum degeneracy point approach [D. Vion \emph{et al.}, Science \textbf{296}, 886 (2002)] effectively protects superconducting qubits from low-frequency noise that couples with the qubits as transverse noise. However, low-frequency noise in superconducting qubits can originate from various mechanisms and can couple with the qubits either as transverse or as longitudinal noise.  Here, we present a quantum circuit containing a universal quantum degeneracy point that protects an encoded qubit from arbitrary low-frequency noise. We further show that universal quantum logic gates can be performed on the encoded qubit with high gate fidelity. The proposed scheme is robust against small parameter spreads due to fabrication errors in the superconducting qubits. 
\end{abstract}


\maketitle

\section{Introduction\label{Sec Introduction}}
Decoherence due to the low-frequency noise is commonly considered as the major hurdle for implementing fault-tolerant quantum computing in superconducting qubits \cite{SchonReview, YouReview, ClarkeReview}. The low-frequency noise, often with $1/f$-type of spectrum \cite{DuttaHorn,Weissman}, is ubiquitous in Josephson junction devices \cite{NECCharge1f, FluxQubitnoise, PhaseQubit1fCC, YYu2008}. In the past, extensive efforts have been devoted to study the microscopic origin of the low-frequency noise \cite{Zorin, PhaseQubit1fTemp, Charge1fTemp, SQUIDDephasing, Koch, CCYU, FluxGeometry}. Most recently,  theoretical and experimental researches suggested that one source of the low-frequency noise is the spurious two-level system fluctuators in the substrate or in the oxide layers of the Josephson junctions \cite{Shnirman, PhaseQubitTLS2, TLSAPL, TLSNature, CCYUTLS}.

To protect the coherence of the superconducting qubits from the low-frequency noise, various approaches have been developed during the past few years, including the dynamic control technique, the quantum degeneracy point approach, the calibration of the qubit parameters by continuous measurement, and the designing of novel quantum circuits and materials \cite{NECEcho, SlowPulseDD, Compensation, DDT, DD1f,  Vion,IthierPRB, MakhlinOptimal, TianPRL2007, PhaseQubitTLS2, transmon1, transmon2, fluxonium, material}. Among these approaches, the quantum degeneracy point approach \cite{Vion,IthierPRB, MakhlinOptimal} has been demonstrated to protect the qubit effectively from the low-frequency noise that couples with the qubit through the off-diagonal matrix elements, i.e. the transverse noise.   The qubit decoherence time was increased by orders of magnitude by operating the qubit at its quantum degeneracy point, also called the optimal point or the ``sweet spot'', where the first order derivative of the qubit energy to the noise fluctuation is zero. This approach has already been applied to both the charge qubit and the flux qubit \cite{SchrieflThesis, NECFluxQubitq1f}.

Meanwhile, due to the diverse origins of the low-frequency noise in solid-state systems, the qubit can couple with either transverse or longitudinal low-frequency noise, where the longitudinal noise couples with the qubit in the diagonal elements and shifts the qubit energy. The simple quantum degeneracy point approach can only protect the qubit from transverse low-frequency noise. For longitudinal low-frequency noise, this approach can't reduce the decoherence of the qubit.  For instance, in the quantronium qubit \cite{Vion}, when the low-frequency noise appears as a fluctuation in the Josephson energy, the qubit energy obtains an energy shift proportional to the noise, which generates qubit dephasing. 

In this work, we propose a universal quantum degeneracy point (UQDP) scheme where the encoded qubits can be protected from generic low-frequency noise.  The physical qubits in this scheme are subject to transverse and (or) longitudinal low-frequency noises. We construct encoded qubit in a subspace where the low-frequency noise only generates off-diagonal elements and can be effectively treated as transverse noise.  We will show that universal quantum logic gates can be implemented in this architecture and are protected from the low-frequency noise as well. To test the analytical results, we numerically simulate the quantum logic gates in the presence of the low-frequency noise. The gate operations, protected by the encoding, demonstrate high fidelity in the simulation.  Moreover, we will show that the proposed scheme is robust again small fabrication errors in the parameters of the Josephson junctions. The paper is organized as follows. In Sec. \ref{Sec Universal point}, we first present the UQDP scheme and the formation of the encoded qubits. Then, the decoherence of the encoded qubits under generic low-frequency noise is calculated analytically.  In Sec. \ref{Sec Gates}, we study the realization of the quantum logic gates on the encoded qubits.  Numerical simulation of the quantum logic gates is also presented in this section. The discussions on the effects of the parameter spreads, different choices of the coupling between the physical qubits, and comparison with the Decoherence Free Subspace (DFS) approach are presented in Sec. \ref{Sec Conclusion} together with the conclusions.

\section{Universal Quantum Degeneracy Point (UQDP) \label{Sec Universal point}}
The basic idea of the (simple) quantum degeneracy point approach is to use the finite energy separation between the two eigenstates of a qubit to protect the qubit from transverse low-frequency noise. Consider the qubit coupling with the transverse noise as
\begin{equation}
H_{\mathrm{s}}=E_{z}\sigma _{z}+\delta V_{x}(t)\sigma
_{x}  
\label{Equa Single Sample}
\end{equation}
where the energy separation between the qubit states $|\uparrow\rangle$ and $|\downarrow\rangle$ is $2 E_{z}$  and $\delta V_{x}(t)$ is the low-frequency noise with $ |\delta V_{x}(t)| \ll E_{z}$. The noise couples with the qubit via the $\sigma_x$ operator which provides a transverse coupling in the off-diagonal elements.  Here, we treat $\delta V_{x}(t)$ as a  classical noise for simplicity, but our results can be applied to quantum noises. The low-frequency nature of the noise determines that it can't resonantly (effectively) excite the qubit between its two states due to the large energy separation between the qubit states. Hence, the noise can be treated as static fluctuations. The qubit energy can be written as 
\begin{equation}
H_{\mathrm{s}}\approx (E_{z}+\delta V_{x}^{2}(t)/2E_{z})\sigma _{z}
\label{Eq:approxHs}
\end{equation}
by second order perturbation approach, i. e. the qubit Hamiltonian adiabatically follows the time dependence of the noise via the second order term $\delta V_{x}^{2}(t)/2E_{z}$. The qubit dephasing is determined by this second order term and is hence significantly suppressed by a factor of $\sim \left\vert \delta V_{x}(t)/2E_{z}\right\vert^2$.

However, as we mentioned in Sec. \ref{Sec Introduction}, in real experiments, the qubit-noise coupling can be more complicated than that in Eq. \ref{Equa Single Sample}. In this work, we consider a generic noise model with $\sum \delta V_{\alpha}(t)\sigma _{\alpha }$ including an arbitrary coupling with the qubits in all Pauli operators. We will show that an encoded subspace can be constructed in which the generic low-frequency noise can be converted to a transverse noise for the encoded qubit.

\subsection{The encoded qubit \label{SubSec Logic Qubit}}
The encoded qubit can be constructed from two identical superconducting qubits connected by a coupling circuit.  The general form of the total Hamiltonian for the coupled qubits can be written as 
\begin{equation}
H_{\mathrm{0}}=E_{z}(\sigma _{z1}+\sigma _{z2})+\sum_\alpha E_{m\alpha} \sigma_{\alpha1}\sigma _{\alpha2} \label{Eq Coupled Qubit Bare}
\end{equation}
where $\sigma _{\alpha j}$ are the Pauli operators of the $j$-th qubit and $\alpha=x,y,z$.   It can be shown that couplings in this general form can generate the encoded qubit under the condition: $E_{mx}\ne 0$ and (or)  $E_{my}\ne 0$. In the following, we set $E_{my}=E_{mz}=0$ with a finite $E_{mx}$ for simplicity of discussion.  The low-frequency noise coupled with the qubits have the general form 
\begin{equation}
V_{\mathrm{n}}=\sum_{\alpha j} \delta V_{\alpha j}(t)\sigma _{\alpha j} \label{Eq:Vn}
\end{equation}
where $\delta V_{\alpha j}(t)$ accounts for the noise coupling with $\sigma_{\alpha j}$ of the $j$-th qubit. The total Hamiltonian including the system and the noise can be written as $H_{\mathrm{en}} = H_{\mathrm{0}}+V_{\mathrm{n}}$.

The coupling Hamiltonian in Eq.(\ref{Eq Coupled Qubit Bare}) can be realized in many circuits.   For example, consider two charge qubits biased in their quantum degeneracy points with effective Josephson energy $E_{J}$, Josephson capacitance $C_{J}$, and gate capacitance $C_{g}$. We then have $E_{z}=E_{J}/2$. The qubits are connected by a superconducting quantum interference device (SQUID) with capacitance $C_{m}$ and tJosephson energy $E_{J2}$ \cite{SchonReview}. It can be derived that 
\begin{equation}
E_{mx}=C_{m}e^{2}/[\left( C_{m}+C_{J}+C_{g}\right) ^{2}-C_{m}^{2}]\label{eq:Em}
\end{equation}
and $E_{my}=E_{mz}=-E_{\mathrm{J2}}/4$. Similar couplings can be derived for other superconducting qubits such as phase qubits and flux qubits.

The eigenstates of the Hamiltonian $H_{\mathrm{0}}$ in Eq. (\ref{Eq Coupled Qubit Bare}) can be derived as
\begin{equation}
\begin{array}{l}
|1\rangle =-\sin \theta |\downarrow\downarrow\rangle +\cos \theta |\uparrow\uparrow\rangle,  \\ 
|2\rangle =\cos \theta |\downarrow\downarrow\rangle +\sin \theta |\uparrow\uparrow\rangle,  \\ 
|3\rangle =\left( -|\downarrow\uparrow\rangle +|\uparrow\downarrow\rangle \right) /\sqrt{2},  \\
|4\rangle =\left( |\downarrow\uparrow\rangle +|\uparrow\downarrow\rangle \right) /\sqrt{2}, 
\end{array}
\label{Eq Eigen}
\end{equation}
with 
\begin{equation}
\cos\theta=2E_z/\sqrt{4E_z^2+E_m^2}\label{eq:cos}
\end{equation}
and $\theta\in [0,\pi/2]$. The corresponding eigenenergies are $E_1=-\sqrt{4E_z^2+E_m^2}$, $E_2=\sqrt{4E_z^2+E_m^2}$, $E_3=-E_m$, and $E_4=E_m$ respectively.  We can rewrite the Pauli operators of the physical qubits in the eigenbasis in the order from $|1\rangle$ to $|4\rangle$. For example, in the eigenbasis, we have
\begin{equation}
\sigma _{z1}=\left[
\begin{array}{cccc}
-\cos \theta & -\sin \theta & 0 & 0 \\
-\sin \theta & \cos \theta & 0 & 0 \\
0 & 0 & 0 & -1 \\
0 & 0 & -1 & 0
\end{array}
\right] ,  \label{Eq Z1}
\end{equation}
and 
\begin{equation}
\sigma _{y2}=i\left[
\begin{array}{cccc}
0 & 0 & \cos \phi & \sin \phi \\
0 & 0 & \sin \phi & -\cos \phi \\
-\cos \phi & -\sin \phi & 0 & 0 \\
-\sin \phi & \cos \phi & 0 & 0
\end{array}
\right] .  \label{Eq Y2}
\end{equation}
with the angle $\phi=\theta/2+\pi/4$. 

An interesting observation is that the diagonal elements of all the Pauli matrices in the subspace of  the states $\left\{ |3\rangle, |4\rangle \right\} $ are zero, i.e.
\begin{equation}
\left\langle 3\right\vert \sigma _{\alpha j}\left\vert 3\right\rangle
=\left\langle 4\right\vert \sigma _{\alpha j}\left\vert 4\right\rangle
=0
\end{equation}
for $\alpha =x,y,z$ and $j=1,2$. The only non-zero matrix elements in this subspace are the off-diagonal elements $\langle 3|\sigma_{zj}|4\rangle$ and their conjugate elements.  We know that the qubits couple with the low-frequency noise through the Pauli matrices as is given in Eq.(\ref{Eq:Vn}). Hence, the noise only couples with this subspace through off-diagonal coupling elements. This coupling, however, will not generate effective excitation between the states $|3\rangle$ and $|4\rangle$ due to the low-frequency nature of the noise and the finite energy difference $2 E_{m}$ between these two states.  We thus select $\left\{ |3\rangle, |4\rangle \right\}$ as the subspace for the encoded qubit and name the parameter point where the Hamiltonian has the form of Eq. (\ref{Eq Coupled Qubit Bare}) as the ``universal quantum degeneracy point'' (UQDP). Note that because of the generality of the form of the Hamiltonian, the UQDP may not be just one point, but can be a curve in the parameter space.  At the UQDP, the subspace $\left\{ |3\rangle ,|4\rangle \right\} $ couples with all the noise $\delta V_{\alpha j}(t)$ transversely and suffers only quadratic dephasing. In addition, due to the same reasons that protect the encoded qubit from dephasing, the leakage from the encoded subspace to the redundant space of $\left\{ |1\rangle, |2\rangle \right\} $ due to the perturbation of the noisy is also prohibited in the lowest order. The matrix elements of $\sigma _{xj}$ and $\sigma _{yj}$ (hence the noise components $\delta V_{x j}(t)$ and $\delta V_{y j}(t)$ induce only virtual excitations between $\left\{ |3\rangle ,|4\rangle \right\} $ and $\left\{ |1\rangle ,|2\rangle \right\}$. Note we only consider decoherence due to the low-frequency noise which is the dominant source of decoherence for superconducting qubits.  High-frequency noises such as Johnson noise in the electric circuits can generate transitions between the encoded subspace and the redundant space.   

\subsection{Dephasing of the encoded qubit}
The dephasing of the encoded qubit can be calculated with perturbation theory. Without loss of generality, we assume that the noise contains only the $x$ component $\delta V_{xj}(t)$ and the $z$ component $\delta V_{zj}(t)$, both of which are Gaussian $1/f$ noises with the spectra
\begin{equation}
S_{xj}(\omega )=\int_{-\infty }^{\infty }\left\langle \delta V_{xj}(t)\delta V_{xj}(0)\right\rangle \frac{e^{i\omega t}dt}{2\pi }=\frac{A^{2}\cos ^{2}\eta}{\omega}\label{Sx}
\end{equation}
\begin{equation}
S_{zj}(\omega )=\int_{-\infty }^{\infty }\left\langle \delta V_{zj}(t)\delta V_{zj}(0)\right\rangle \frac{e^{i\omega t}dt}{2\pi }=\frac{A^{2}\sin ^{2}\eta}{\omega}\label{Sz}
\end{equation}
where the angle $\eta$ is a parameter that describes the noise power distribution and $A^2$ is the total noise power. When $\eta=0$, the low-frequency noise is a transverse noise with only the $x$ (and $y$) component; When $\eta=\pi/2$, the low-frequency noise is a longitudinal noise with only the $z$ component.

We introduce the Pauli operators for the encoded qubit as $X=|3\rangle \left\langle 4\right\vert +|4\rangle \left\langle 3\right\vert $, $Y=-i|3\rangle \left\langle 4\right\vert +i|4\rangle \left\langle 3\right\vert $, and $Z=|3\rangle \left\langle 3\right\vert -|4\rangle \left\langle 4\right\vert $.  When projected to the encoded subspace of $\left\{ |3\rangle ,|4\rangle \right\} $, the effective Hamiltonian of the system coupling with the low-frequency noise can be written as
\begin{equation}
H_{\mathrm{en}}= \left(-E_{m}-\frac{E_{m}(\sum \delta V_{xj}(t))^{2}}{2E_{z}^{2}} + \frac{(\sum \delta V_{zj}(t))^{2}}{2E_{m}}\right)Z \label{eq:Hen}
\end{equation}
by applying second order perturbation.  The noise enters this Hamiltonian in quadratic form.  The dephasing of the encoded qubit due to arbitrary low-frequency noise hence only involves quadratic terms, similar to Eq.(\ref{Eq:approxHs}) for the simple quantum degeneracy point.  Even the longitudinal noise $\delta V_{zj}$ only contributes to the dephasing in quadratic terms. 

\begin{figure}
\includegraphics[width=7.5cm,clip]{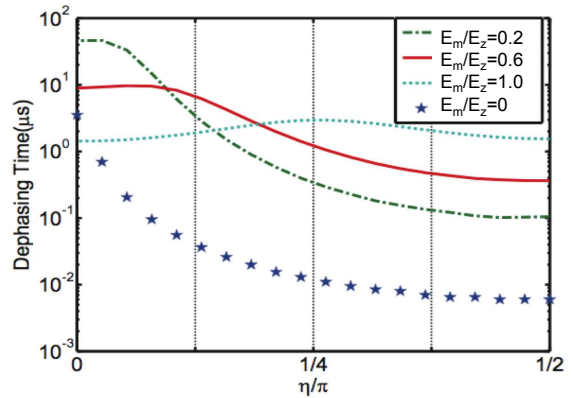} 
\caption{Dephasing time of the bare qubit (star) and the encoded qubit at different ratio $E_{m}/E_{z}$ as is labelled in the plot (other curves). }
\label{Fig FIDLaw}
\end{figure}

To calculate the dephasing of the encoded qubit, we use the analytical results on dephasing due to Gaussian $1/f$ noise \cite{MakhlinOptimal}. In the calculation, we assume the noises $\delta V_{\alpha j}(t)$ are not correlated with each other \cite{HuQuadratic}. The parameters we choose are $E_{z}/2\pi\hbar =5\,\mathrm{GHz}$, $A=2 \times 10^{-4}E_{z}$ with an infrared cutoff for the $1/f$ noise $\omega _{ir}/2\pi =1\, \mathrm{Hz}$.  The dephasing times are plotted in Fig.~\ref{Fig FIDLaw} at different coupling ratio $E_m/E_z$.  For the bare qubit, the decoherence time decreases rapidly by a few orders of magnitude as the noise distribution changes from transverse noise to longitudinal noise.  For the encoded qubit at $E_m/E_z=1$, in contrast, the  decoherence time only varies smoothly over the whole range of $\eta$. 

\section{Protected Quantum Logic Gates for Encoded Qubits\label{Sec Gates}}
In the previous section, we showed that the encoded qubit is immune to arbitrary low-frequency noise to the first order of the coupling and the dephasing is dominated by quadratic terms derived from perturbation theory. As a result, the encoded qubit can be a highly-coherent quantum memory for storing quantum information. In this section, we will further show that universal quantum logic gates on the encoded qubits are also protected from the low-frequency noise with high fidelity. The gate operations require manipulations on the physical qubits.  We will test our theoretical results for the gate operations with a numerical simulation.  The numerical results give high gate fidelity for the UQDP.

\subsection{Single-qubit gates}
Quantum logic gates on single encoded qubit can be performed by manipulating the operators of individual physical qubits $\sigma_{\alpha i}$ or by manipulating the interaction terms between two physical qubits $\sigma_{\alpha i}\sigma_{\alpha j}$.  In the encoded space, the effective Hamiltonian of the encoded qubit is $H_{\mathrm{0}}^{\mathrm{(e)}}=-E_m Z$.  When projected to this subspace, the operators $\sigma _{z1}$ can be expressed as
\begin{equation}
P_{e}\sigma _{z1}P_{e}=-X \label{Eq Z XXZ}
\end{equation}
where $P_{e}$ is the projection operator onto the subspace $\left\{ |3\rangle, |4\rangle \right\}$. From Eq. (\ref{Eq Z1}), it can also be shown that the operator $\sigma _{z1}$ generates no coupling between the encoded subspace and states in $\left\{ |1\rangle, |2\rangle \right\}$. By pumping the first physical qubit with a pulse $H_{\mathrm{X}}=2\lambda \cos (2E_{m}t/\hbar)\sigma _{z1}$ for a duration of $ \theta \hbar/2\lambda $, we can implement the $X$-rotation gate $U_{x}(\theta)=\exp (i\theta X/2)$. 

Similarly, the operator $\sigma _{y1}\sigma _{y2}+\sigma _{z1}\sigma_{z2}$ can be expressed as
\begin{equation}
P_{e}(\sigma _{y1}\sigma _{y2}+\sigma _{z1}\sigma _{z2})P_{e}=-1- Z \label{Eq SQUID XXZ}
\end{equation}
which generates a rotation in the $Z$-component of the encoded qubit. This operator can be realized in superconducting circuits. For example, for charge qubits connected by a coupling SQUID, this operator can be realized by varying the flux in the SQUID loop. By applying an ac pumping on the coupling SQUID with $H_{\mathrm{Z}}=2\lambda \cos (2E_{m}t/\hbar)(\sigma _{y1}\sigma _{y2}+\sigma _{z1}\sigma _{z2})$ for a duration of $\theta \hbar/2\lambda $, we have a  $Z$-rotation gate $U_{z}(\theta )=\exp(i\theta Z/2)$. Combining the operations in Eq. (\ref{Eq Z XXZ}) and Eq. (\ref{Eq SQUID XXZ}), we achieve a complete $SU(2)$ generator set that generates arbitrary single-qubit quantum logic gates on the encoded qubit. 

\subsection{Controlled quantum logic gates}
\begin{figure}
\includegraphics[width=8cm,clip]{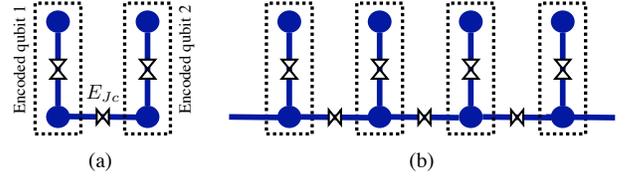}
\caption{(a) Schematic plot of two encoded qubits coupling via a tunable circuit. The physical qubits are represented by the solid circles and are connected by the vertical bars to form the encoded qubits. The lower physical qubits in each encoded qubit are connected to achieve the coupling between the encoded qubits. (b) An array of encoded qubits connected with their nearest-neighbors.}
\label{Fig Coupled Logic Qubit}
\end{figure}

Two-qubit gates on the encoded qubits can be achieved by connecting the circuits of the encoded qubits as is shown in Fig.~\ref{Fig Coupled Logic Qubit}a. We consider two encoded qubits and use $\sigma_{\alpha j}$ as the Pauli operators for the first encoded qubit and $\tau_{\alpha j}$ as Pauli operators for the second encoded qubit. Assume that the lower physical qubits in each encoded qubit are connected with the coupling Hamiltonian
\begin{equation}
H_{\mathrm{cgate}}=- 2 \lambda _{c}\cos (2(E_{m1}-E_{m2})t/\hbar) \sigma _{z2}\tau _{z2}.
\label{Eq:cgate}
\end{equation}
The operators can be projected as
\begin{equation}
P_{e}\sigma _{z2}P_{e}=X_1 \label{Psz1}
\end{equation}
\begin{equation}
P_{e}\tau _{z2}P_{e}=X_2\label{Psz2}
\end{equation}
onto the encoded subspaces.  Hence, in the rotating frame the above coupling Hamiltonian can be written as
\begin{equation}
H_{\mathrm{cgate}}^{\mathrm{(rot)}}=-\frac{\lambda _{c}}{2}(X_{1}X_{2}+Y_{1}Y_{2}),
\end{equation}
which provides a swap-like interaction between the encoded qubits. Combining this interaction with the single-qubit gates $U_{x}(\theta)$ and $U_{z}(\theta)$, we obtain a universal set of quantum logic operations for the encoded qubits. We can also design various geometries to connect arrays of encoded qubits. For example, in Fig.~\ref{Fig Coupled Logic Qubit}b, the lower physical qubits of the encoded qubits are connected to their nearest neighbors in a chain.  In this case, the system can be viewed as a one-dimensional chain of qubits with nearest neighbor coupling. Universal quantum computation can be implemented in this configuration.

For superconducting qubits, the above coupling Hamiltonian can be constructed using a tunable Josephson junction (a SQUID) that connects the physical qubits.  For a SQUID with Josephson energy $E_{Jc}$ and capacitive energy $E_{cc}$, we have
\begin{equation}
H_{\mathrm{cgate}}=-\frac{E_{Jc}}{4}(\sigma _{z2}\tau _{z2}+\sigma_{y2}\tau _{y2})+E_{cc}\sigma _{x2}\tau _{x2}
\label{Eq LogicQubitCouplingXXZ}
\end{equation}
where both the Josephson energy and the capacitive energy can be tunable \cite{YouJianQiangTunablePRL, tunableCC, Sembat1, tunableCCLtian}.  The coupling includes the extra terms $\sigma_{y2}\tau _{y2}$ and $\sigma _{x2}\tau _{x2}$, which only contain off-resonant leakage terms such as $|1\rangle_\sigma \langle 3| \otimes |3\rangle_\tau \langle 1|$.  Under the time-dependent modulation $\cos(\omega_d t)$ by choosing proper pumping frequency, these terms can be neglected. 

\begin{figure}
\includegraphics[width=7.5cm,clip]{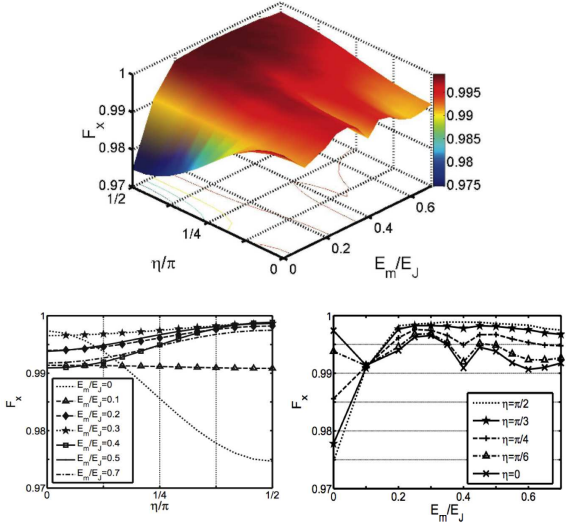}
\caption{Gate fidelity $F_{x}$ of single-qubit gate $U_{x}(\pi)$ versus $\eta$ and $E_{m}$. Here,  $E_{J}=2E_{z}$. Bottom left, $F_{x}$ versus $E_{m}$ at various $\eta$ as is labelled in the plot.  Bottom right, $F_{x}$ versus $\eta$ at various $E_{m}/E_{z}$ as is labelled in the plot.}
\label{FX3D}
\end{figure}
\subsection{Numerical simulation}
To test the analytical results in the previous subsections, we numerically simulate the quantum logic gates. In the simulation, we made the following assumptions on the $1/f$ noise: 1. all the physical qubits couple with the environmental noise in the form of $\delta V_{xj}(t)\sigma_{xj}+\delta V_{zj}(t)\sigma_{zj}$;  2. the noise power $A^2$ and the noise distribution angle $\eta$ are the same for all physical qubits.  The simulation of the $1/f$ noise can be implemented as
\begin{equation}
V_{\alpha j} (t)=\sum_{\omega =\omega _{ir}}^{\omega_{uv}} a_{\alpha j}(\omega) \cos (\omega t+\phi ) \Delta\omega\label{Vsimu}
\end{equation}
where $a_{\alpha j}$ is a gaussian distribution with zero average and satisfies 
\begin{equation}
\langle a_{\alpha x}(\omega_1)a_{\alpha x}(\omega_2)\rangle=A^2\cos^2(\eta) \delta(\omega_1+\omega_2)/\omega_1\label{aax}
\end{equation}
\begin{equation}
\langle a_{\alpha z}(\omega_1)a_{\alpha z}(\omega_2)\rangle=A^2\sin^2(\eta) \delta(\omega_1+\omega_2)/\omega_1\label{aaz}
\end{equation}
in connection with the definitions in Eq.(\ref{Sx}) and Eq.(\ref{Sz}). Here, discrete noise components are used to replace the continuous integral on the spectral density with $\Delta\omega/2\pi=10^{-4}\,\textrm{MHz}$. The phase $\phi$ is a random number with the uniform distribution between $0$ and $2\pi$. The parameters we choose for the physical qubits and the noise are $E_{z}/2\pi\hbar =5\,\mathrm{GHz}$ for the physical qubits, the infrared limit of the noise frequency $\omega _{ir}/2\pi =1\,\mathrm{Hz}$, the upper bound of the noise frequency $\omega_{uv}/2\pi =0.1\,\mathrm{MHz}$,  and the noise power $A/E_{z}=2\times10^{-4}\,\mathrm{s/rad}$.

For the single-qubit operation $U_{x}(\pi )=\exp (i\pi X/2)$, which generates a bit-flip gate for the encoded qubit, we use the following definition for the gate fidelity \cite{NielsonFidelity} 
\begin{equation}
F_x=\frac{1}{2}+\frac{1}{12}\sum_{i=1,2,3} \mathrm{Tr}(U_{x}(\pi )\Sigma _{i}U^{\dag}_{x}(\pi )\varepsilon (\Sigma_{i}))\label{FX_def}
\end{equation}
where $\Sigma_i=X,\,Y,\,Z$ for $i=1,2,3$ are the Pauli operators for the encoded qubit and $\varepsilon (\Sigma_{i})=P_{e}\mathcal{L}(\Sigma_{i})P_{e}^\dag$ is the projection of the final state onto the encoded subspace after applying the quantum process on the initial density matrix $\Sigma_{i}$.  The coupling constant is chosen to be $\lambda /2\pi\hbar =300\,\mathrm{MHz}$. The gate fidelity $F_{x}$ of $U_{x}(\pi)$ versus $\eta$ and $E_{m}$ is plotted in Fig.~\ref{FX3D}. For finite coupling $E_m$, where the encoded subspace is protected by the coupling, the fidelity only varies smoothly as the noise varies from transverse noise to longitudinal noise.  While for $E_m=0$, i.e. the uncoupled (bare) qubits, the fidelity decreases sharply as $\eta$ increases to $\pi/2$.  At fixed noise distribution $\eta$, the fidelity first rises rapidly when the coupling $E_{m}$ increases from zero, then becomes saturated and even shows oscillatory behavior as $E_{m}$ further increases. With $E_m \sim  0.4E_{z}$, $F_{x}$ exceeds $0.997$ for $\eta \in \lbrack 0,\pi /2]$. In this regime, the optimal $E_{m}$ for a particular $\eta$ usually can be found within the range of  $(0.4 E_{z},\,E_{z})$. 

\begin{figure}
\includegraphics[width=6.5cm,clip]{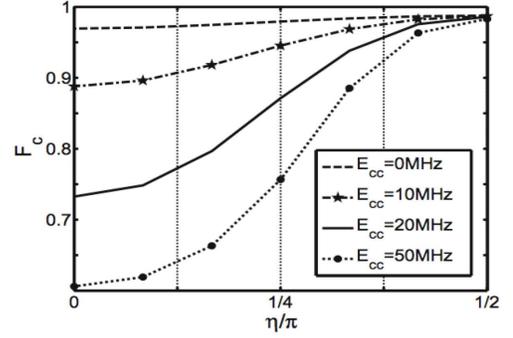} 
\caption{The gate fidelity $F_{C}$ of two-qubit gate $U_{C}$ versus $\eta$ at various $E_{cc}/2\pi\hbar$ as is labelled in the plot.}
\label{Fig FC}
\end{figure}

For the two-qubit operation $U_{C}=\exp [-i\pi (X_{1}X_{2}+Y_{1}Y_{2})/4]$, the gate fidelity can be defined as \cite{NielsonFidelity} 
\begin{equation}
F_C=\frac{1}{5}+\frac{1}{80}\sum_{i,j} \mathrm{Tr} (U_{C}(\Sigma _{i}\otimes\Omega_{j})U^{\dag}_{C}\varepsilon (\Sigma_{i}\otimes\Omega_{j}))\label{FC_def}
\end{equation}
where $\Omega_i=X,\,Y,\,Z$ for $i=1,2,3$ are the Pauli operators for the second encoded qubit and the super-operator gives $\varepsilon (\Sigma_{i}\otimes\Omega_{j})=P_{e}\mathcal{L}(\Sigma_{i}\otimes\Omega_{j})P_{e}^\dag$. We choose the coupling of the first encoded qubit to be $E_{m1}/2\pi\hbar =5\,\mathrm{GHz}$ and the coupling of the second encoded qubit to be $E_{m2}/2\pi\hbar =2\,\mathrm{GHz}$. The operating Hamiltonian is given by Eq. (\ref{Eq LogicQubitCouplingXXZ}) and is applied for a duration of $\pi \hbar /4 \lambda _{c}$ with $\lambda _{c}/2\pi\hbar =300\,\mathrm{MHz}$.  The gate fidelity $F_{c}$ for $U_{C}$ versus $\eta$ is plotted in Fig. \ref{Fig FC} at various capacitive coupling $E_{cc}$. For zero $E_{cc}$, the gate $U_{C}$ can be accomplished with high fidelity and shows only weak dependence on the distribution angle $\eta$, which proves the ``universality'' of our scheme. However, for non-zero $E_{cc}$, the fidelity drops quickly, and hence the implementation of the two-qubit gate requires the capacitive coupling to be small. 

\subsection{State preparation and detection}
To implement the above quantum logic gates, we need to prepare the initial state of system in the encoded subspace. This can be achieved by letting the coupled qubits relax to their ground states $|1\rangle$ via thermalization. Then, an ac driving
\begin{equation}
H_{\mathrm{prep}}=2\lambda _{p}\cos \left[ (E_{1}-E_{m})t/\hbar\right] \sigma _{x1}
\end{equation}
can be applied  on the first physical qubit,  which generates a Rabi oscillation between the states $|1\rangle$ and $|3\rangle$ given the non-zero matrix element $\langle 3| \sigma _{x1}|1\rangle =-\sin(\theta/2+\pi/4)$. After a period of $\pi /2\lambda _{p}\sin(\theta/2+\pi/4)$, the state becomes $|3\rangle$ which is now in the encoded subspace.

State detection of the encoded qubits can be implemented with the assistance of single-qubit rotations as well. To measure the probability of an encoded qubit in the state $|3\rangle$ or $|4\rangle$, apply single-qubit gate $\exp [-i\pi X/4]$ as discussed above which converts the state $|3\rangle$ to the state $|\uparrow\downarrow\rangle$ and converts the state $|4\rangle $ to the state $|\downarrow\uparrow\rangle $. Then, measuring the physical qubits can give us information of the encoded states.

\section{Discussions and Conclusions \label{Sec Conclusion}}
The encoded qubit proposed above is made of two identical physical qubits with qubit energy $E_{z}$.  For superconducting qubits, this parameter can be the Josephson energy $E_{J}$ of a charge qubit in the degeneracy point or the energy gap in the flux qubit. This parameter depends on the parameters of the Josephson junctions in the circuit. In real devices, the junction parameters usually have fabrication errors with an error spread on the order of $5\%$.  Below we discuss the effect of the parameter error on the proposed scheme. We introduce a factor $a_{0}$ as the ratio between the energy of the second qubit and the energy of the first qubit with $|a_0-1|\ll 1$.  The Hamiltonian in Eq. (\ref{Eq Coupled Qubit Bare}) becomes
\begin{equation}
H_{\mathrm{0n}}=E_{z}(\sigma _{z1}+a_{0}\sigma _{z2})+E_{m} \sigma_{z1}\sigma _{z2}. 
\label{Eq NonUniformHami}
\end{equation}
The eigenbasis of this Hamiltonian still includes the encoded subspace $\left\{ |3\rangle ,|4\rangle \right\}$, but with the new eigenstates and eigenenergies
\begin{equation}
\begin{array}{lcl}
|3\rangle _{\mathrm{n}}=\cos \varphi |3\rangle +\sin \varphi |4\rangle , &
\quad & E_{\mathrm{3n}}=-E_{\mathrm{m,n}}; \\
|4\rangle _{\mathrm{n}}=-\sin \varphi |3\rangle +\cos \varphi |4\rangle , &
\quad & E_{\mathrm{4n}}=E_{\mathrm{m,n}};
\end{array}
\end{equation}
where $E_{\mathrm{m,n}}=\sqrt{(1-a_{0})^{2}E_{\mathrm{J}}^{2}+E_{\mathrm{m}}^{2}}$ and 
\begin{equation}
\varphi =1/2\sin ^{-1}[(a_{0}-1)E_{\mathrm{z}}/E_{\mathrm{m,n}}]\label{varphi}
\end{equation}
respectively. The operators $\sigma _{x1}$, $\sigma _{x2}$, $\sigma_{y1}$, and $\sigma _{y2}$ only contain non-vanishing matrix elements connecting $\left\{ |1\rangle ,|2\rangle \right\}$ and $\left\{ |3\rangle ,|4\rangle \right\}$. However, $\sigma _{z1}$ and $\sigma _{z2}$ now include non-trivial diagonal terms 
\begin{equation}
\begin{array}{l}
_{\mathrm{n}}\langle 3|\sigma _{z1}|3\rangle _{\mathrm{n}} = -_{\mathrm{n}}\langle 4|\sigma _{z1}|4\rangle _{\mathrm{n}}=\frac{-(1-a_{0})E_{\mathrm{z}}
}{E_{\mathrm{m,n}}}, \\
_{\mathrm{n}}\langle 3|\sigma _{z2}|3\rangle _{\mathrm{n}} = -_{\mathrm{n}}\langle 4|\sigma _{z2}|4\rangle _{\mathrm{n}}=\frac{(1-a_{0})E_{\mathrm{z}}
}{E_{\mathrm{m,n}}},  
\end{array}\label{extra34}
\end{equation}
which induce residual longitudinal noises on the encoded qubit. The residual coupling can be derived as
\begin{equation}
V_{\mathrm{res}}\approx \frac{(a_{0}-1)E_{\mathrm{z}}}{E_{\mathrm{m,n}}} (\delta V_{z1}-\delta V_{z2})Z \label{Eq ResidualCoupling}
\end{equation}
in terms of the effective Pauli operator $Z$, which generates dephasing in first order terms of $\delta V_{zj}$.  However, the residual coupling contains the ratio $|a_0-1|$ of the error spread  which is less than $5\%$ or even smaller with the advance of current technology. The dephasing by this term is hence reduced by a factor of $|a_0-1|^2< 2\times10^{-3}$.  Given the recent experimental measurement \cite{IthierPRB} of  $|\delta V_{zj}/E_{\mathrm{z}}| <  10^{-2}$, the longitudinal dephasing due to the error spread can be comparable with or even lower than the quadratic dephasing due to the transverse noise in Eq. (\ref{eq:Hen}).

We would like to mention that the proposed scheme is different from the Decoherence Free Subspace (DFS) approach that has been widely studied in quantum information processing \cite{LidarDFS}.  The DFS approach protects qubits from spatially correlated noises by choosing a subspace that is immune to such noises, i.e. the dephasing is suppressed by the noise correlation.  While in our scheme, we explore the energy separation of the encoded subspace and the low-frequency nature of the noise (which can't excite transitions between states with large energy separation in the first order) to protect the quantum coherence. 

In Sec. \ref{Sec Universal point} and \ref{Sec Gates}, we studied the encoded subspace and the gate operations using the coupling Hamiltonian with $E_{mx}=E_{m},\, E_{my}=E_{mz}=0$ for the simplicity of our discussions. It can be shown that the UQDP approach can be applied to the general form of coupling given in Eq. (\ref{Eq Coupled Qubit Bare}) as far as either $E_{mx}\ne 0$ or $E_{my}\ne 0$ can be satisfied. As an example, consider the situation  
\begin{equation}
E_{mx}=E_{m},\, E_{my}=E_{mz}=b_0 E_m\label{eq:H02}
\end{equation}
with a coefficient $b_0$. In real system, this coupling can be obtained from a Josephson junction with energy $E_{J2}$ that connects the two physical qubits and $b_{0}=-\frac{E_{J2}}{4E_{m}}$. It can be found that the states $|3,4\rangle$ defined in Eq. (\ref{Eq Eigen}) still form the subspace for the encoded qubit. The energies of these states become $\epsilon_{3} = -E_{m}-2b_{0}E_{m}$ and $\epsilon_{4} = E_{m}$ including a shift due to the finite $b_0$.  It can also be shown that the projections to the encoded subspace are $-P_{e}\sigma_{z1}P_{e}=P_{e}\sigma_{z2}P_{e}=X$ and $P_{e}\sigma_{xi}P_{e}=P_{e}\sigma_{yi}P_{e}=0$. The encoded qubit is protected against any first order dephasing by the low-frequency noise. This observation shows that the UQDP scheme can be applied to various superconducting qubits such as flux qubits and phase qubits as far as we can construct the coupling in Eq.(\ref{Eq Coupled Qubit Bare}). 

In conclusion, we have proposed a scheme of a universal quantum degeneracy point (UQDP) that can protect the superconducting qubits from generic low-frequency noise.  Using coupled qubits to form the encoded qubits, we find a subspace where the low-frequency noise only affects the qubit dephasing to quadratic order. We have shown that universal quantum logic gates can also be performed on the encoded qubits with high fidelity.  The scheme is robust again parameter spreads due to fabrication errors. The scheme can be applied to systems with very general form of couplings and provides a promising approach to protect superconducting qubits against low-frequency noise.

\section{Acknowledgments}
This work is supported by the National Science Foundation under Grant No. NSF-CCF-0916303 and NSF-DMR-0956064. XHD is partially supported by Scholarship from China.

\end{document}